\documentclass[aps,twocolumn]{revtex4}
\usepackage{epsf}
\usepackage{graphicx}
\usepackage{amsmath}

\renewcommand{\baselinestretch}{1.}
\newcommand{\dir}{Figs}

\newcommand{\eq}{ {Eq.} }
\newcommand{\fig}{ {Fig.} }
\newcommand{\figs}{ {Figs.} }
\newcommand{\etal}{ {\em et al.} }
\newcommand{\ie}{ {\em i.~e.}, }
\newcommand{\eg}{ {\em e.~g.}, }

\begin{document}
\setlength{\parskip}{1.3ex}

\def\twiddle{\lower.9ex\rlap{$\kern-.1em\scriptstyle\sim$}}
\def\bigtwiddle{\lower1.ex\rlap{$\sim$}}
\def\gtwid{\mathrel{\raise.3ex\hbox{$>$\kern-.75em\lower1ex\hbox{$\sim$}}}}
\def\ltwid{\mathrel{\raise.3ex\hbox{$<$\kern-.75em\lower1ex\hbox{$\sim$}}}}
\def\half{{1 \over 2}}
\def\third{{1 \over 3}}
\def\tthird{{2 \over 3}}
\def\fourth{{1 \over 4}}
\def\fthrd{{4 \over 3}}
\def\fivehalf{{5 \over 2}}
\def\kt{k_B T}
\def\bR{ {\bf R} }
\def\br{ {\bf r}}
%
%
%

\title{Formation and structure of the microemulsion phase in two-dimensional
ternary AB+A+B polymeric emulsions}
\author{Dominik D\"uchs and Friederike Schmid}

\affiliation{Fakult\"at f\"ur Physik, Universit\"at Bielefeld, \\
Universit\"atsstr. 25, 33615 Bielefeld, Germany}

\begin{abstract}
We present an analysis of the structure of the fluctuation-induced microemulsion 
phase in a ternary blend of balanced AB diblock copolymers with equal amounts of 
A and B homopolymers. To this end, graphical analysis methods are employed to 
characterize two-dimensional configuration snapshots obtained with the recently 
introduced Field-Theoretic Monte Carlo (FTMC) method. We find that a microemulsion 
forms when the mean curvature diameter of the lamellar phase coincides roughly 
with the periodicity of the lamellar phase. Further, we provide evidence to the 
effect of a subclassification of the microemulsion into a genuine and a 
defect-driven region.
\end{abstract}

\maketitle

\section{INTRODUCTION}
Microemulsions arise when two immiscible substances are compatibilized by the
use of a surfactant lowering the interfacial tension between the incompatible
components dramatically. Originally introduced for mixtures of oil, water, and
a surfactant, the term microemulsion applies equally to polymeric blends. 
Microemulsions appear in two forms: droplet and bicontinuous. 
Here we consider the bicontinuous case, which is observed in symmetric
systems with balanced surfactants. In polymeric emulsions, the compatibilizing 
copolymer is usually the most expensive ingredient. A good theoretical 
understanding of these system is therefore desirable from an application point of 
view, as well. Applications often depend crucially on the morphology of the blend, 
and bicontinuity is a particularly useful property, \eg for improved stiffness 
or conductivity.

Traditionally, there have been three distinct coarse-grain approaches to the 
theoretical study of self-assembling amphiphilic systems\cite{gompper}: 
(a) {\it Microscopic} approaches\cite{rajagopalan}, often built around lattice-gas or
lattice-Boltzmann simulations\cite{love1,sakai,theissen,nekovee,chen,love2};
also by dissipative particle dynamics\cite{jury,shillcock,guo}
or by standard Monte Carlo or molecular dynamics 
methods\cite{karaborni,marcus,bearchell,bedrov,mark,poncela,loison,bourov}.
(b) Ginzburg-Landau models\cite{gompper1,cruz,holyst,schwarz,clarysse,tanja},
leading to a much simpler description by means of only a few order parameters
and coefficients which can be obtained from experiment. 
Polymeric blends have been described very successfully with a particular
continuous density functional approach usually denoted self-consistent field
theory (SCFT)\cite{edwards,helfand,schmid}. 
The drawback of this method is that it employs a
mean-field approximation and therefore neglects the very cause for the
existence of microemulsions in these systems. 
(c) Membrane approaches for systems in which the solubility of the amphiphiles 
is extremely low, such that they can form membranes by themselves\cite{gompper2}. 
This case is not considered here.

As mentioned above, the addition of a compatibilizing agent, here the AB
copolymer, is essential to the formation of mesoscopically ordered phases in a
pure A and B homopolymer mixture. The melts would otherwise phase-separate
macroscopically below some relatively high temperature, which is typically
above room temperature. Microemulsions, although by definition part of the
disordered regime, do have a finite (mesoscopic) preferential length scale, and
thus their existence depends on the presence of a compatibilizer, as
well. Apart from the microscopic correlation length, $\xi$, which is the
characteristic {\em decay} length of the correlation function, a second
length scale, $q^{-1}$, which describes the wavelength of {\em oscillations} 
in the correlation function, comes into play.  The disorder line is now
defined as the locus where $q$ just vanishes. The onset of the microemulsion
regime, on the other hand, is defined by the Lifshitz line, which is the locus 
where the {\it peak} in the structure factor is just found at nonzero wavevector 
$q^*$. A microemulsion is thus disordered but not structureless\cite{teubner,gompper}.

In a recent paper, we introduced a new approach to incorporating the effect of
thermal fluctuations in field theories of polymer melts: the Field-Theoretic
Monte Carlo (FTMC) method\cite{duechs}. It is an extension of the earlier mentioned
SCFT method. The system under investigation was the ternary model system consisting
of symmetric AB copolymers as well as A and B homopolymers, the homopolymers
being $\alpha = 0.2$ times as long as the copolymers. This system had previously 
been the object of a series of experiments conducted by 
Bates \etal \cite{bates1,morkved1,hillmyer}. Mean-field
calculations of its phase diagram predict a Lifshitz critical point where the
disordered, lamellar, and phase-separated regions meet\cite{LP,diehl}. 
However, it could be demonstrated both in these experiments as well as in the 
FTMC simulations that the Lifshitz point is destroyed by thermal fluctuations
and a channel of bicontinuous microemulsion ($B\mu E$) emerges in between the
lamellar and phase-separated regions. Whereas the aim of that paper, 
Ref.~\onlinecite{duechs}, was mainly to establish the validity of the FTMC method 
and to compare it with a related complex Langevin method\cite{venkat,glenn1,glenn2},
we here want to examine more closely the formation and structure of the 
microemulsion phase. 

Neither the disorder line nor the total monomer Lifshitz line correspond to any
thermodynamic transition. Recently, Morkved \etal's \cite{morkved2} have found that
neither line correlates well with the transition from a fully disordered
mixture at higher temperatures to a well-developed $B\mu E$. Nevertheless, 
they found that dynamic light scattering provided a clear signal for this
transition, which they proposed to be the homopolymer/homopolymer Lifshitz
line originally introduced by Holyst and Schick\cite{holyst1}. In a similar 
experimental system, Schwahn \etal \cite{schwahn} found evidence for the existence 
of {\em three} different regimes in the disordered phase, a ``disordered blend'', 
a ``microemulsion'', and a ``disordered copolymer'' region. The three regimes 
differ from each other by the value of the peak wavevector $q^*$ in the structure 
factor -- zero in the disordered blend, large in the disordered copolymer, 
intermediate in the microemulsion.
On the theoretical side, Holyst and Przybylski\cite{holyst2} performed Monte 
Carlo simulations of a Ginzburg-Landau model for the lamellar phase in copolymers 
and showed that topological fluctuations change the monotonic decay of the 
off-specular scattering intensity, introducing the {\it topological Lifshitz line}. 
One might suspect that such fluctuations affect the characteristic lamellar distance 
to higher values. Experiments\cite{bates2} as well as simulations\cite{fried} 
indeed reveal that the peak wavevector $q^*$ is $\sim 15-20 \%$ lower at the 
order-disorder transition than predicted by SCFT. However, this effect can also 
be explained by local chain stretching\cite{fried,maurer}.

With our approach, we can separate the effect of fluctuations from that of local 
chain rearrangements. It combines the advantages of SCFT, which accounts
in full for the chain connectivity but neglects fluctuations, with those of 
Ginzburg-Landau simulations, which include fluctuations but make approximations
for the intrachain correlations. Our results are qualitatively similar to those of
Morkved \etal and Schwahn \etal Our main result in this paper is the 
subclassification of the microemulsion region into a {\it disordered} 
and a {\it genuine} regime. For the transition from the lamellar to this 
disordered microemulsion phase we do not find a markedly lowered $q^{*}$. 
In the genuine regime, $q^*$ differs from the mean field value. Indeed, this is 
our criterion for the distinction of the two. It should be interesting to examine 
to what extent, if any, our subclassification corresponds to 
Morkved \etal's \cite{morkved2} distinction of poor and good microemulsions, 
or to Schwahn \etal's distinction of disordered copolymers and microemulsions.

In the present work, we have used parallel Cray architectures to simulate
AB+A+B melts on two-dimensional lattices of size 48$\times$48. The high demands
on computing power have so far limited us to two dimensions. It will
be straightforward to adapt the method to three dimensions once this becomes
feasible from a computational point of view. This paper is organized as
follows. In Section II, we shall briefly outline the theoretical model
underlying FTMC. For a more detailed description, we refer the interested
reader to our earlier paper\cite{duechs}. Section III contains the results and 
discussion. In this context, our evidence is presented to the effect of a
subclassification of the microemulsion. We conclude in Section IV with a
summary. 

\section{THE FIELD-THEORETIC MODEL} 
\label{sec2}
In this section, we briefly present the model underlying FTMC for the ternary
system under consideration. For details, see Ref.~\onlinecite{duechs}. 
We study a mixture of $n_A$ homopolymers of type A, $n_B$ homopolymers of type B, 
and $n_{AB}$ symmetric block copolymers in a volume $V$. The polymerization index of 
the copolymer is denoted by $N$, and the corresponding quantities for the
homopolymers are denoted as by $N_A = N_B = \alpha N$. We consider the case of
a symmetric copolymer, wherein the fraction of A monomers in the copolymer is $f
= 1/2$. We restrict our attention to the concentration isopleth, where the
homopolymers have equal volume fractions $\phi_{HA}=\phi_{HB}=\phi_{H}/2$. The
monomeric volumes of both A and B segments are assumed to be identically equal
to $1/\rho_0$. On this model we impose an incompressibility constraint. We
model the effective interactions between segments as Flory-Huggins local
contact interactions and use a Gaussian chain model, which corresponds to
perfectly flexible polymers. With these assumptions, the canonical partition
function of the system can be written as 
\begin{equation}
\label{eq8}
{\cal Z}_C \propto \int_{\infty} D W_- \int_{i\infty} D W_+ \exp [ - H_C (W_-, W_+)]
\end{equation}
with
\begin{eqnarray}
\label{eq9}
\lefteqn{
H_C (W_-, W_+) =
C \Big[\frac{1}{\chi N} \int d{\br} \, W_-^2 - \int d{\br} \, W_+  
} \\&&
- V(1-\phi_H) \ln Q_{AB} 
- \frac{V\phi_H}{2\alpha} \ln Q_A 
- \frac{V\phi_H}{2\alpha} \ln Q_B \Big],
\nonumber
\end{eqnarray}
\begin{equation}
C=\frac{\rho_0}{N} R^d_{g}.
\end{equation}
In this paper, all lengths are expressed in units of the
unperturbed radius of gyration, $R_{g}=b(N/(2d))^{1/2}$, where $d$ is the
space dimension. The parameter $C$ in the above equations, which occurs
as a global prefactor to $H_C$, acts as a Ginzburg parameter such 
that in the limit 
$C\rightarrow\infty$ the partition function (\ref{eq8}) is reduced to 
its saddle point and the mean-field solution becomes exact. In \eq~(\ref{eq9}),
$Q_A, Q_B, {\text{ and }} Q_{AB}$ denote the single chain partition functions 
for the A, B, and AB chains, respectively, in the potential fields 
$W_- (\br) {\text{ and }}W_+ (\br)$. Note that $W_-$ is conjugate 
to the difference in A and B densities, $\hat m$, and
$W_+$ to the total density, $\hat\phi$. Moreover, $W_-$ is real, whereas $W_+$
is imaginary thereby rendering $H_C$ complex. We here employ a partial
saddle point approximation in $W_+$, which reduces $H_C$ to real values.

The single chain partition functions can be expressed in terms of the
Feynman-Kac formulae \cite{helfand} as:
\begin{equation}
\label{eq13}
Q_i = \int d \br \; q_i (\br ,\nu_i) ,
\end{equation}
where the propagators $q_i$ satisfy diffusion equations. From these
propagators, we can calculate
density operators, $\bar\phi_A$ and $\bar\phi_B$, from
\begin{eqnarray}
\label{eq19}
\bar{\phi}_A (\br) &=  &
\frac{V(1-\phi_H)}{Q_{AB}} \int_0^f \!\!\! ds \; q_{AB}
(\br ,s) q^\dagger_{AB} ( \br ,1-s) 
\\ &&
+ \frac{V\phi_H}{2 \alpha Q_{A}} \int_0^\alpha ds
\; q_A (\br ,s) q_{A} ( \br ,\alpha-s),
\nonumber
\end{eqnarray}
and a similar equation for the $\bar\phi_B$. The densities which correspond to
the experimentally measurable quantities are the averages over
$\bar\phi_{A,B}$, \ie $\phi_{A,B} = \langle \bar\phi_{A,B} \rangle$. For the calculations
in the present work, however, we have treated $\bar\phi_A$ and $\bar\phi_B$ as
instantaneous densities. We then calculated time averages over parameters
calculated on them. Strictly speaking, the quantities $\bar\phi_{A,B}$ are
visualizations of the spatial distributions of fields. The time averages so obtained 
should nevertheless reflect the essential structural properties of the system.

For a detailed description of the FTMC method used to obtain the configuration
series that serve as the input data in this work, see Ref.~\onlinecite{duechs}.

\section{RESULTS AND DISCUSSION}

\begin{figure}[t]
\vspace*{0.5cm}
\center
\begin{minipage}{3in} 
\epsfxsize= 3in \epsfbox{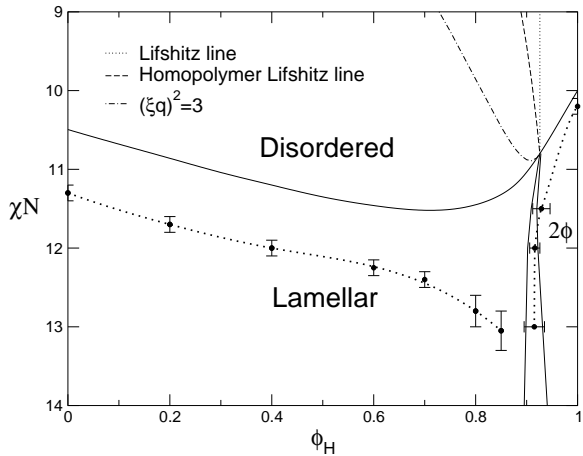}
\end{minipage}
\renewcommand
\baselinestretch{1.}
\caption{Phase diagram of the ternary A+B+AB blend ($\alpha = 0.2$). The solid lines 
show the mean-field phase diagram, which features a three-phase (L+A+B)
coexistence region reaching up to the Lifshitz point. The circles give locations
of fluctuation corrected phase boundaries at $C=50$ (from Ref.~\cite{duechs}).
The thick dotted lines are guides for the eye. The order-disorder transition is
weakly first-order, but the coexistence region is too small to be displayed.
The thin lines in the disordered region indicated the total monomer Lifshitz line 
(dotted), the homopolymer-homopolymer Lifshitz line (dashed), and the line of
$(q \xi)^2=3$ (dashed-dotted), as obtained from mean-field theory.
}
\renewcommand
\baselinestretch{1.5}
\label{C50c}
\end{figure}

\fig~\ref{C50c} shows the fluctuation-corrected phase diagram of the ternary
melt, as obtained in Ref.~\cite{duechs}, at dimensionless polymer number density
$C=50$, which is an intermediate value and has been used throughout this
work. Compared with the corresponding mean-field diagram, one discerns
(a) a shift in the order-disorder transition to higher segregation strengths,
$\chi N$, (b) a slight shift in the same direction of the transition between
the disordered and phase-separated regions, and (c) the emergence of a
cusp-like region of a microemulsion phase. Note that this diagram was
calculated in two dimensions, due to numerical constraints. Likewise, the
simulations carried out in the present work were in two dimensions
only. While a full three-dimensional analysis will likely yield a more narrow
microemulsion ``channel'' than that shown in \fig~\ref{C50c}, the good
qualitative agreement with experiment indicates that essential aspects of the
phase behavior are captured in two dimensions already. Thus it is justified to
proceed with further analysis.  

In addition to these results taken from our earlier paper, Ref.~\onlinecite{duechs}, 
\fig~\ref{C50c} also displays various Lifshitz lines calculated within the
mean-field approximation. The total monomer Lifshitz line is found at the homopolymer 
volume fraction\cite{broseta}
\begin{equation}
\label{ll_total}
\Phi_H = \frac{1}{1 + 2 \alpha^2},
\end{equation}
and is independent of the incompatibility parameter $\chi N$. 
The homopolymer-homopolymer Lifshitz line, on the other hand,
is independent of $\alpha$, and determined by the equation
\begin{equation}
\label{ll_homo}
\chi N = \sqrt{\frac{8}{\Phi_H (1-\Phi_H)}}.
\end{equation}
It was obtained by generalizing the calculation of Holyst and Schick\cite{holyst1} 
for homopolymer/copolymer length ratio $\alpha=1$ to arbitrary $\alpha$. 
Note that in contrast to the case $\alpha = 1$ studied by Holyst and Schick, 
the homopolymer-homopolymer Lifshitz line is rather close to the total monomer 
Lifshitz line at $\alpha = 0.2$. This is because short homopolymers swell
the copolymer blocks, whereas the longer homopolymers in systems with
$\alpha = 1$ are expelled from the copolymer rich regions. 

Finally, \fig~\ref{C50c} also shows the line $(q \xi)^2 = 3$, which distinguishes 
between microemulsions with weak and strong ordering tendency in confined 
geometries\cite{schmid2}.
Here $\xi$ is the correlation length and $q$ the wavevector $q$ of oscillations in 
the correlation function.  The line $(q \xi)^2 = 3$  was calculated by expanding the 
inverse mean-field structure factor $S(q)^{-1}$ up to fourth order in $q$, 
\ie approximating $S(q)$ by the Teubner-Strey form\cite{teubner}, 
\begin{equation}
\label{teubner}
S(q) = \frac{1}{\omega + g q^2 + c q^4}
\end{equation}
and then determining the value of $\chi N$ where $g^2/\omega c = 1$~\cite{schmid2}.
   
\begin{figure}[h]
\center
\begin{minipage}{2.5in}
\hspace*{-0.2in}
\epsfxsize= 2.5in \rotatebox{270}{\epsfbox{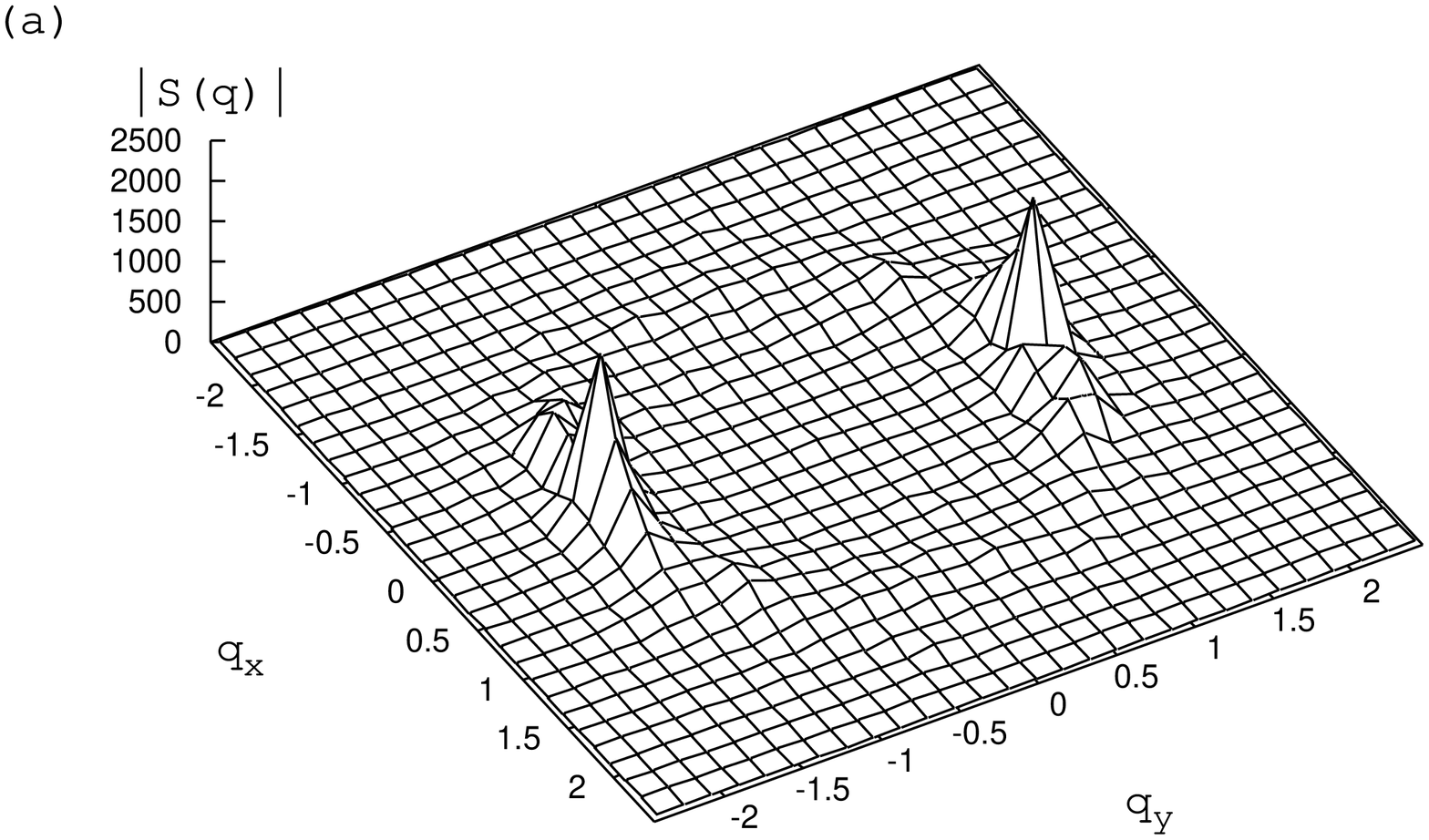}} \\
\hspace*{-0.2in}
\epsfxsize= 2.5in \rotatebox{270}{\epsfbox{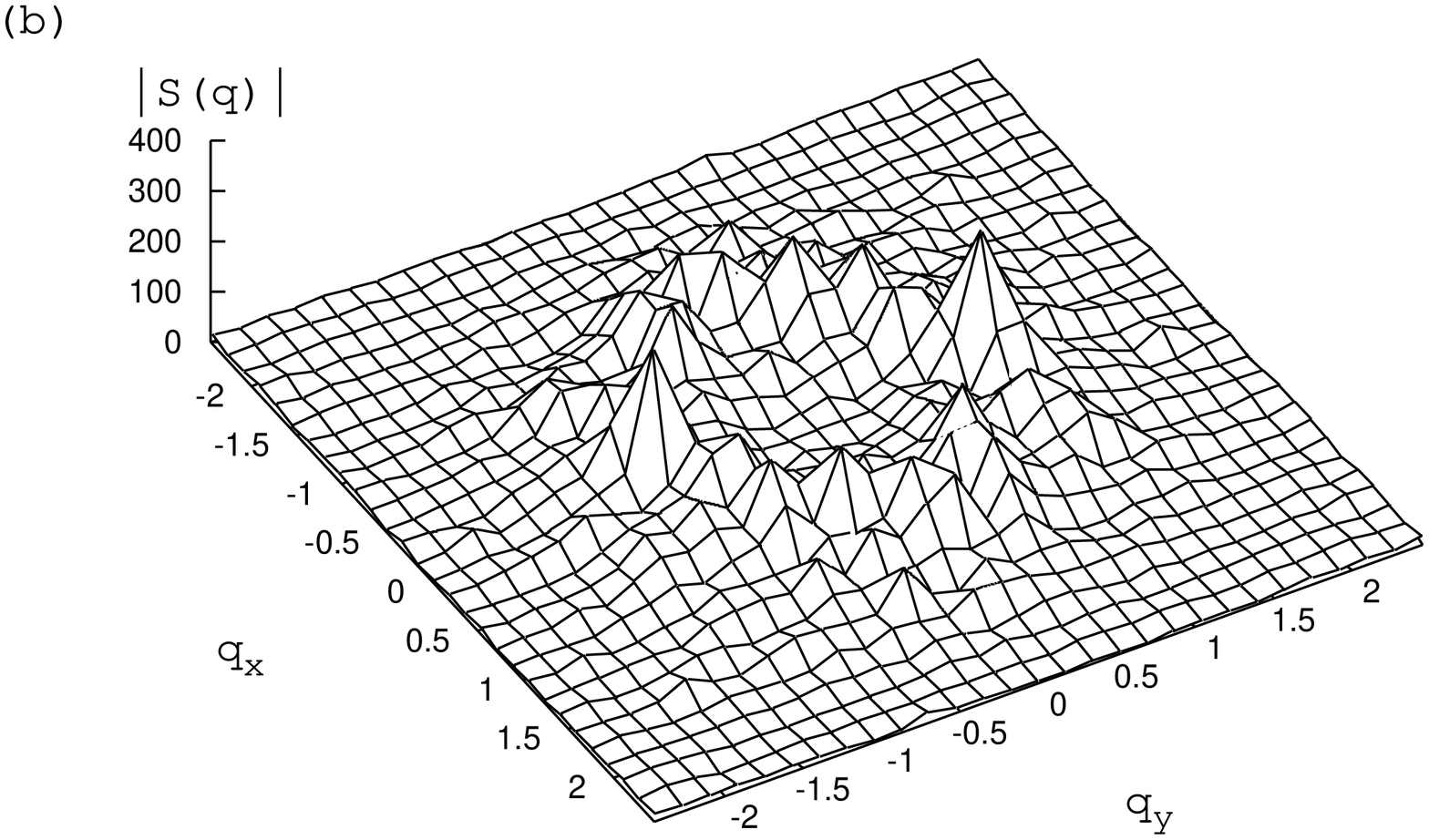}} \\
\hspace*{-0.2in}
\epsfxsize= 2.5in \rotatebox{270}{\epsfbox{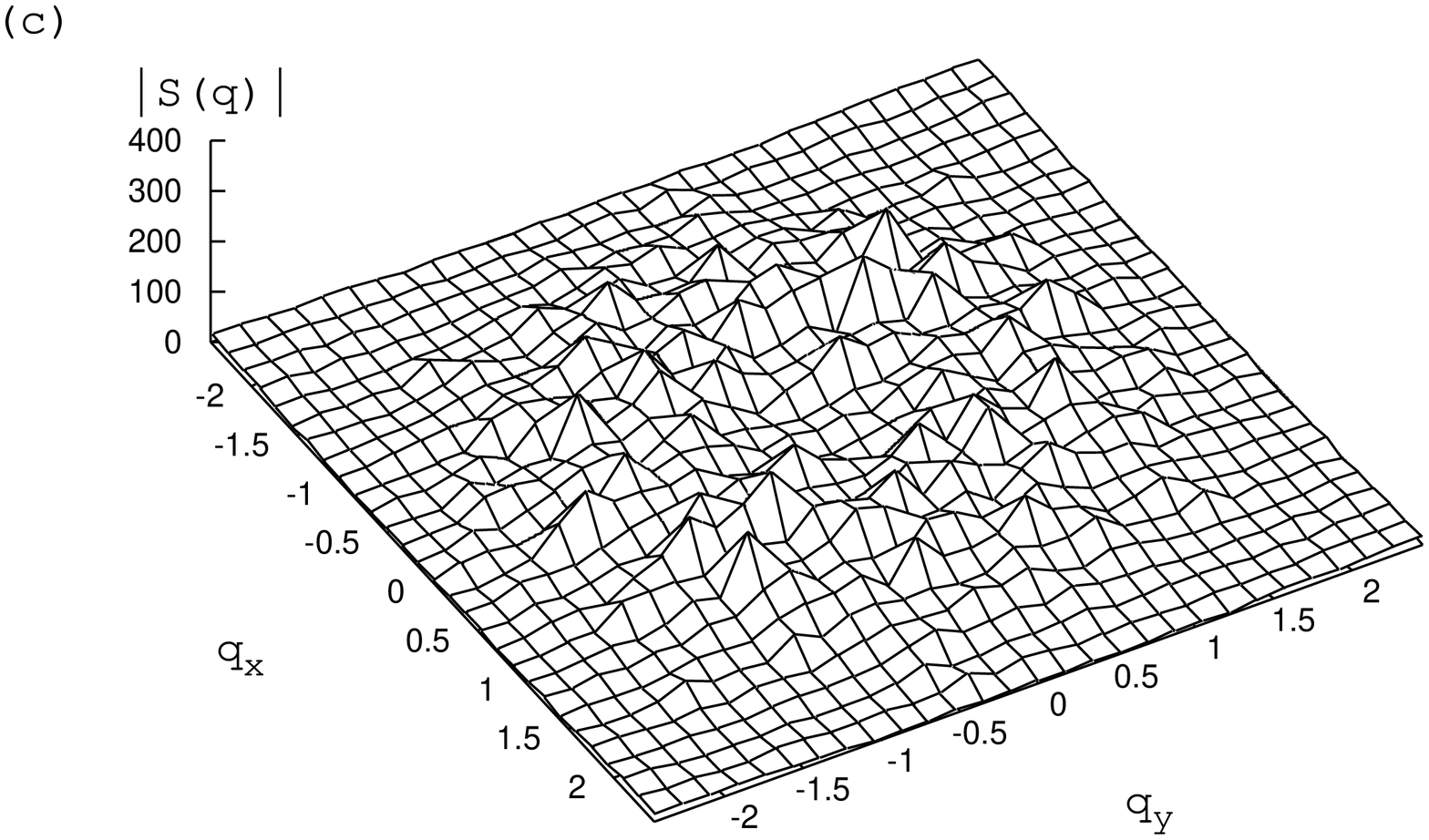}}
\end{minipage}
\renewcommand
\baselinestretch{1.}
\caption{Structure factors at $C=50$
(a) Lamellar phase at $\phi_H=0.7$ and $\chi N = 12.5$.
(b) Strongly structured microemulsion at $\phi_H=0.82$ and $\chi N = 12.5$.
(c) Weakly structured microemulsion at $\phi_H=0.82$ and $\chi N = 10.$
}
\renewcommand
\baselinestretch{1.5}
\label{sfac}
\end{figure}

We start by examining the structure factor of our two-dimensional melts. 
It can be calculated from the Fourier transform of the density correlation
function, which in turn can be obtained from the fluctuating $W_A (\br) \equiv
W_+ (\br) + W_- (\br)$ field via a formula derived in Ref.~\onlinecite{glenn1}. 
Thus, the structure factor $S({\bf q})$ is the Fourier transform of
\begin{equation}
\frac{1}{V}\int\mbox{d}\br_0 
\left[ 
\left\langle\hat\rho(\br_0)\hat\rho(\br_0+\br)\right\rangle -  
\left\langle\hat\rho(\br_0)\right\rangle\left\langle\hat\rho(\br_0+\br)\right\rangle \right],
\end{equation}
with
\begin{eqnarray}
\lefteqn{
\left\langle\hat\rho(\br)\hat\rho(\br')\right\rangle - 
\left\langle\hat\rho(\br)\right\rangle\left\langle\hat\rho(\br')\right\rangle 
 = \frac{2C}{\chi N}\delta(\br-\br') 
} \\ &&
- \frac{4C^2}{(\chi N)^2}
\left[\left\langle W_A(\br)W_A(\br')\right\rangle - \left\langle W_A(\br)\right\rangle
\left\langle W_A(\br')\right\rangle\right].
\nonumber
\end{eqnarray}
In \fig~\ref{sfac}, structure factors for lamellar phases and strongly and weakly 
structured microemulsions are shown. The lamellar structure factor (\fig~\ref{sfac} a) 
features the hallmark double peak of a striped pattern. \fig~\ref{sfac} b) illustrates 
that in a microemulsion, anisotropy is lost yet a preferential length scale does exist, 
as evidenced by a ring-shaped region of maxima. As we progress deeper into the 
disordered phase at lower $\chi N$, the ring becomes less pronounced, until it is almost 
indistinguishable from a a shapeless low-amplitude noise (\fig~\ref{sfac} c).  

The above procedure to calculate the structure factor of a
configuration is quite expensive from a computational standpoint as it involves
integrations over the entire course of a simulation. However, if we are to
analyze geometrical patterns, it is not necessary to use the full-fledged
structure factor. We can see the characteristic features of a configuration
from the Fourier transform of its density distribution, as well. In this work,
we have therefore used time averages of quantities derived from the Fourier
transform of $\bar\phi_A$ in analyzing morphological properties.

\clearpage

%
As a starting point, we need to find the preferential length scale of a density
distribution $\bar\phi_A(\br)$. This is equivalent to finding the preferential
wave vector of its two-dimensional Fourier transform, F({\bf q}). To this end,
we use the square averaged over all angles, defining 
\begin{equation}
F_0(q) := \frac{1}{2\pi}\int_0^{2\pi}d\phi |F({\bf q})|^2,
\label{S0}
\end{equation}
which is now one-dimensional.

\begin{figure}[h]
\vspace*{1cm}
\center
\begin{minipage}{3in}
\epsfxsize= 3in \epsfbox{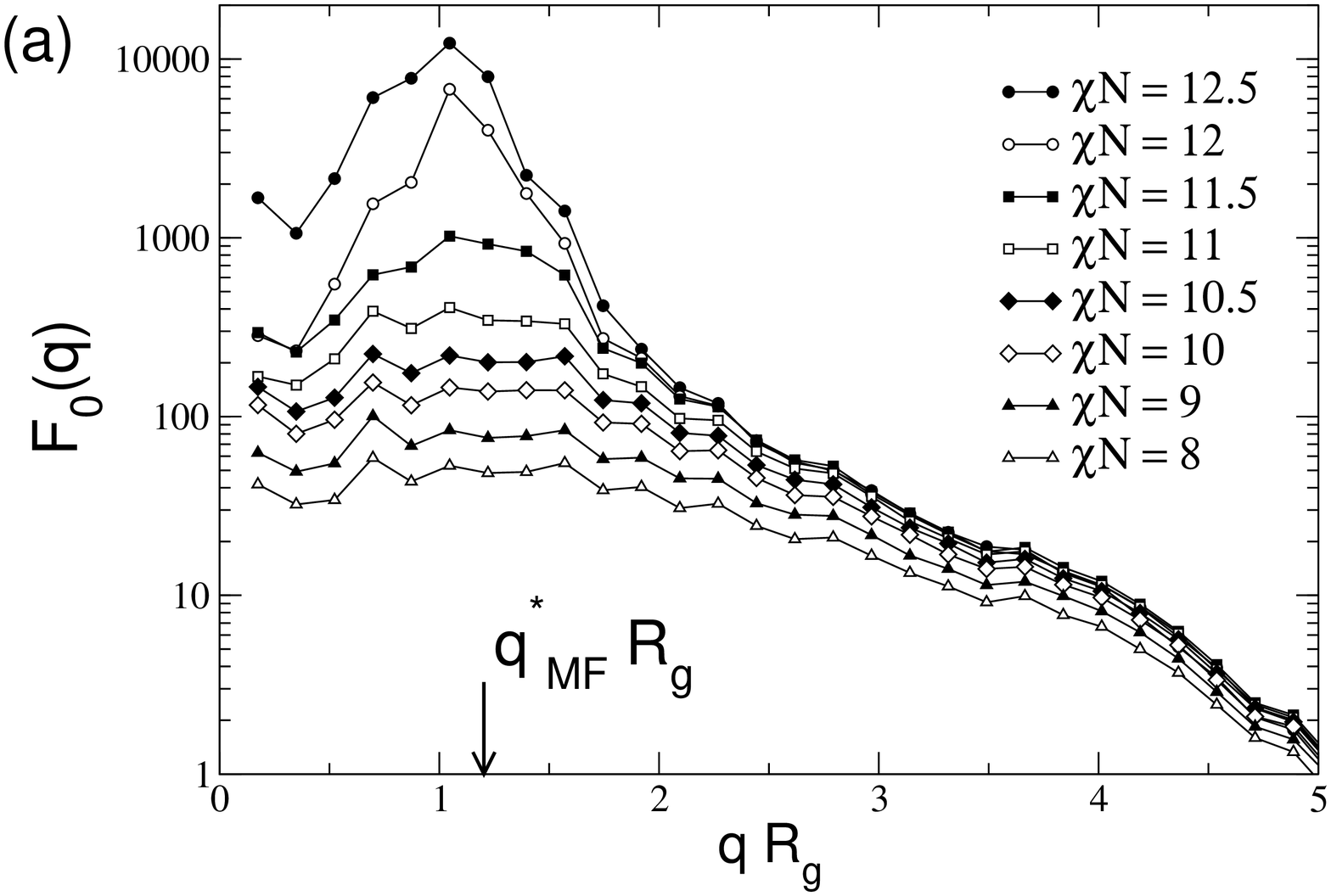}
\epsfxsize= 3in \epsfbox{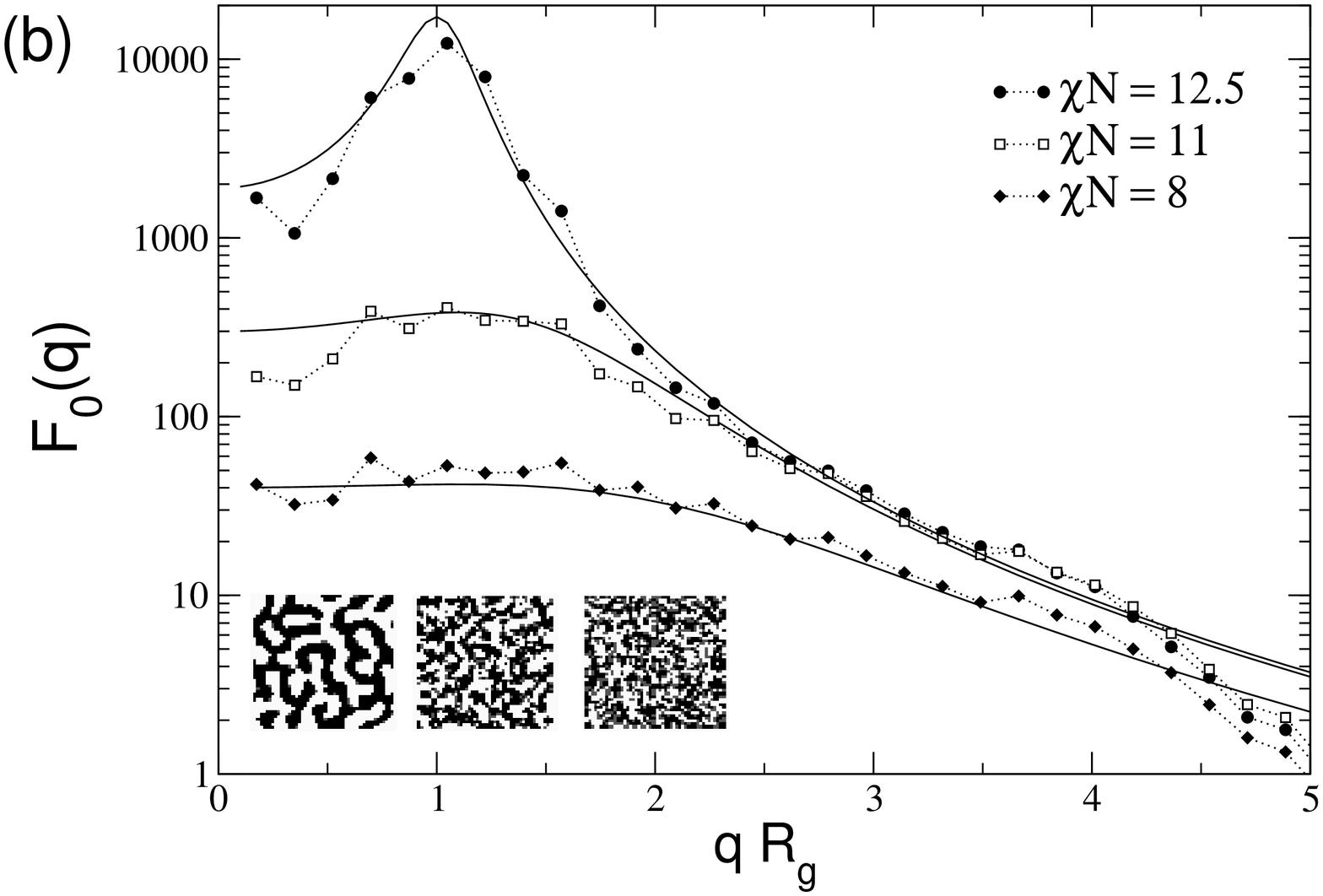}
\end{minipage}
\renewcommand
\baselinestretch{1.}
\caption{
(a) Equilibrated averages of $F_0(q)$ at $C=50$, $\Phi_H = 0.82$, and
different values of $\chi N$. The arrow indicates the mean-field location
$q^*_{MF}$ of the maximum in the structure factor in the disordered phase.
(b) Teubner-Strey fits to $F_0(q)$ for selected values of $\chi N$. 
The inset shows snapshots of $\bar{\phi_A}$ for the same $\chi N$ values 
(12.5, 11, and 8 from left to right).  A lattice point is painted black (``A'') 
if $0.51 \le \bar\phi_A\le 1$, grey if $0.49 \le \bar\phi_A < 0.51$, and
and white (``B'') if $0 \le \bar\phi_A < 0.0.49$.
}
\renewcommand
\baselinestretch{1.5}
\label{logSq}
\end{figure}

From \fig~\ref{logSq}, in which $F_0(q)$ is displayed for various $\chi N$ at
$\phi_H = 0.82$, it can be seen that $F_0(q)$ has a
pronounced peak for $\chi N \stackrel{<}{\sim} 11$. Below $\chi N \sim 11$ the 
curves become markedly flatter at low $q$. Note that the mean-field transition
between the disordered and the lamellar phase occurs at $\chi N = 11.4$ (for
$\Phi_H = 0.82$). Above this point, the position of the (hardly discernible) 
maximum of $F_0(q)$ agrees well with its mean-field value, $q^*_{MF} R_g=1.203$ 
for $\Phi_H = 0.82$. In the more structured region of higher $\chi N$, the peak 
moves to lower values of $q$, reflecting the fact that the corresponding
mean-field length scale, the lamellar distance, increases with increasing
$\chi N$. \fig~\ref{logSq} b) shows fits of selected curves to the Teubner-Strey 
form (\eq~(\ref{teubner}))\cite{teubner}. The fits are reasonable at high and
low values of $\chi N$, and less satisfactory in the vicinity of the mean-field 
transition.

The technicalities of calculating $F_0(q)$ require $q$ to be binned in a
histogram and thus approximated by discrete values (in practice, by multiples
of $2\pi/L$, where L is the box length). In addition, \fig~\ref{logSq} illustrates
that our data for $F_0(q)$ are strongly scattered, mainly due to the statistical
error.  This makes the $q$ coordinate of the maximum of $F_0(q)$ a poor candidate 
to be taken as the preferential wave vector. A smoother procedure is to use 
\begin{equation}
{\bar q} = \frac{\int\mbox{d}q \mbox{ }q \mbox{ }F_0(q)}{\int\mbox{d}q \mbox{ }F_0(q)}.
\label{bar_q}
\end{equation}
If $F_0(q)$ has a pronounced well-defined maximum at nonzero $q^*$ 
(\eg $\chi N \ge 11.5$ in \fig~\ref{logSq}), ${\bar q}$ roughly coincides 
with this maximum. Otherwise, it slightly overestimates $q^*$ 
(\eg $q_0 R_g= 1.3$ at $\chi N = 11.$, $q_0 R_g= 1.6$ at $\chi N = 8$).
Now, the preferential length scale, $L_0$, is defined as 
\begin{equation}
L_0 := \frac{2\pi}{\bar q}.
\label{l0}
\end{equation}

Further, we define the mean curvature diameter, $D_C$, of the boundaries
of A and B microdomains in a black-and-white image similar to those in the
inset of \fig~\ref{logSq} b), in which a lattice point is painted black
(``A'') if $0.5<\bar\phi_A\le 1$, and white (``B'') if 
$0\le\bar\phi_A\le 0.5$:
\begin{equation}
D_C := 2\left[\frac{1}{L_c}\int\mbox{d}s \;  \left|\frac{\mbox{d}{\bf
t}}{\mbox{d}s \; }\right|^2\right]^{-\frac{1}{2}}. 
\end{equation}
$L_c$ is the sum of all contour lengths of the microdomain boundaries,
and {\bf t} is the tangent vector at a given coordinate $s$ along the contour.
For a detailed explanation of the algorithm used to calculate $D_C$, see
Appendix A. Note that the snapshots shown here also contain grey pixels,
corresponding to a balance density, i.e., $0.49<\bar\phi_A<0.51$.

\begin{figure}[h]
\center
\begin{minipage}{2.5in}
\epsfxsize= 2.5in \epsfbox{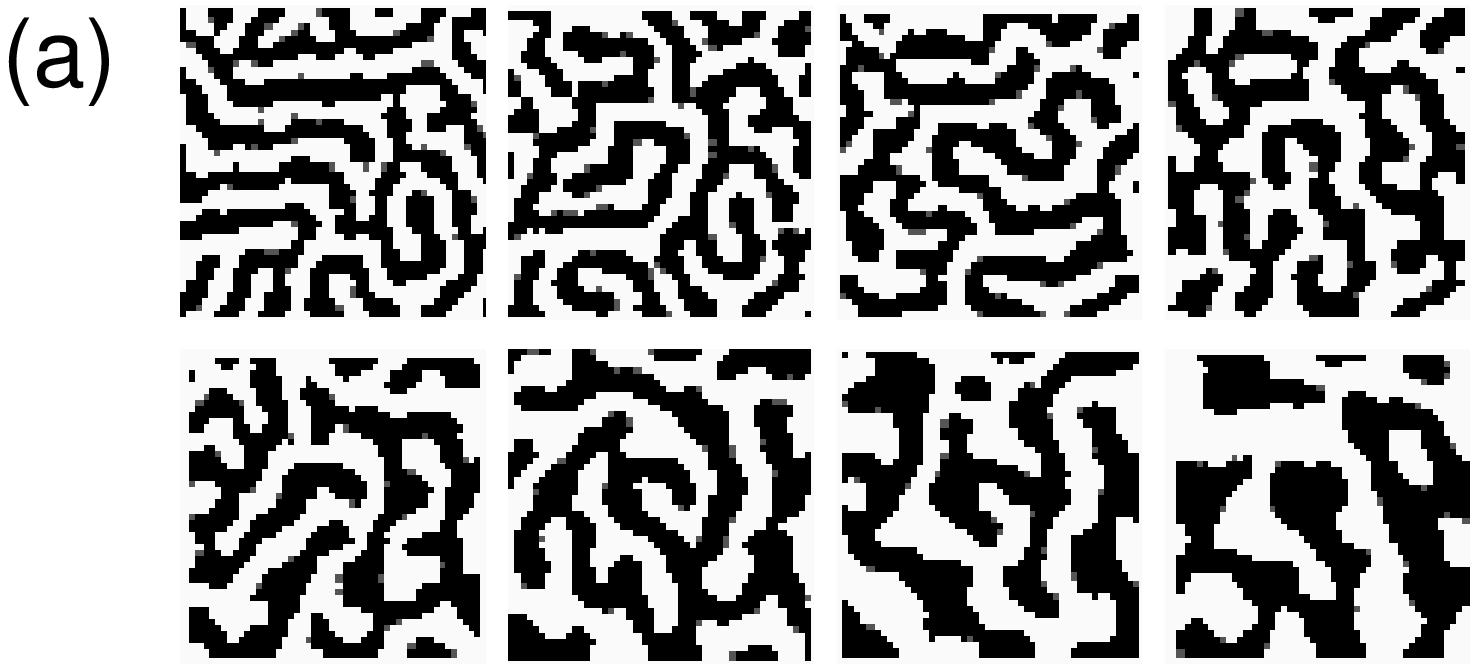}
\vspace*{0.5in}
\epsfxsize= 3in \epsfbox{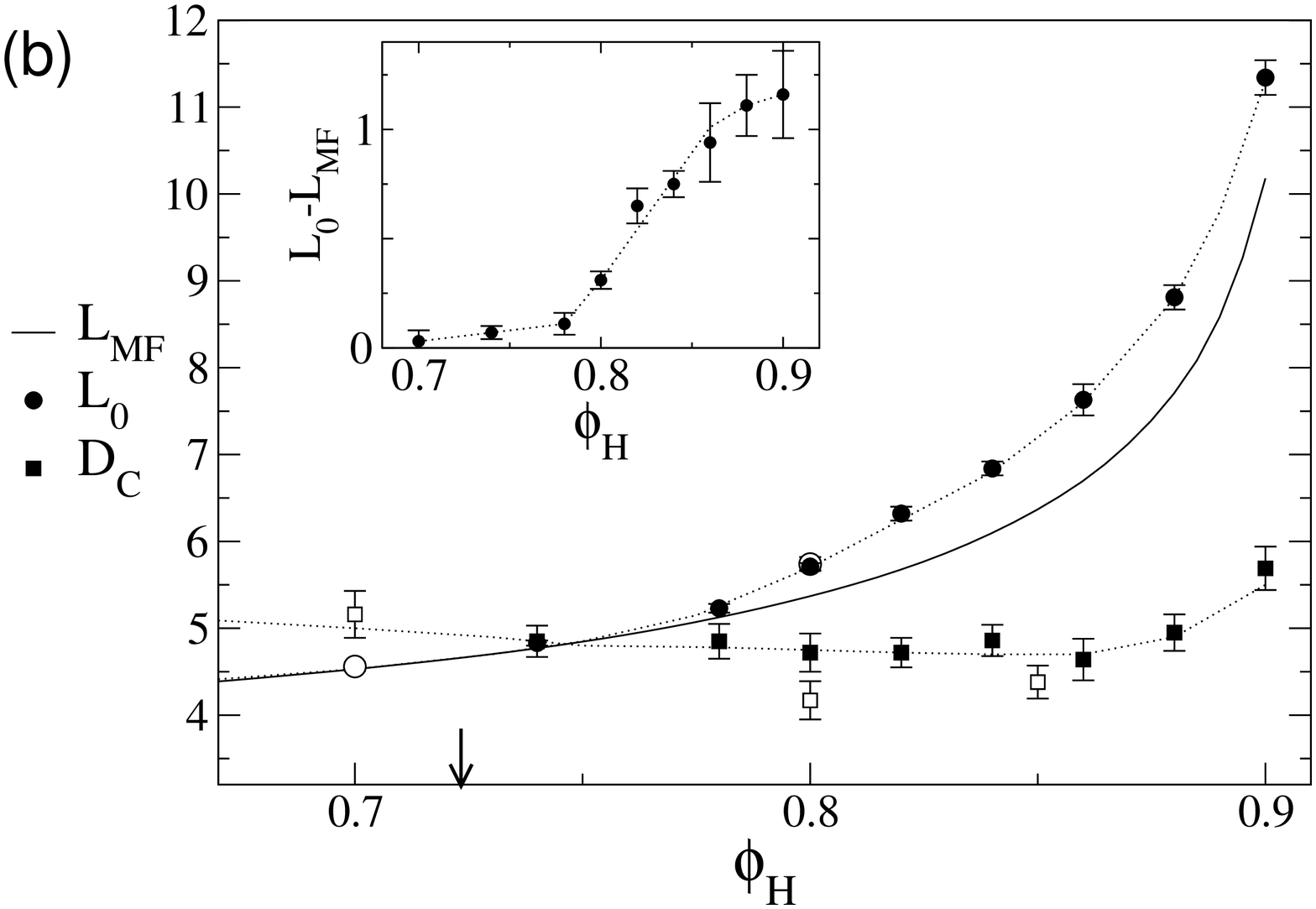}
\end{minipage}
\renewcommand
\baselinestretch{1.}
\caption{
Characteristic length scales at $\chi N = 12.5$, $C=50$.
(a) Snapshots at $\phi_H$ = 0.74, 0.78, 0.8, and 0.82 (first row from left), 
and $\phi_H$ = 0.84, 0.86, 0.88, and 0.9 (second row from left).
(b) Preferential length scale, $L_0$ (circles), and curvature diameter, $D_C$
(squares), in units of $R_g$, vs. $\Phi_H$, averaged over the equilibrated parts 
of simulations on a $48 \times 48$ lattice. The solid line shows the mean-field
lamellar distance $L_{MF}$ for comparison. Filled and empty symbols correspond 
disordered and lamellar initial conditions, respectively. The inset shows the
difference of $L_0$ and $L_{MF}$.
}
\renewcommand
\baselinestretch{1.5}
\label{lengths125}
\end{figure}

\begin{figure}[h]
\center
\begin{minipage}{2.5in}
\epsfxsize= 2.5in \epsfbox{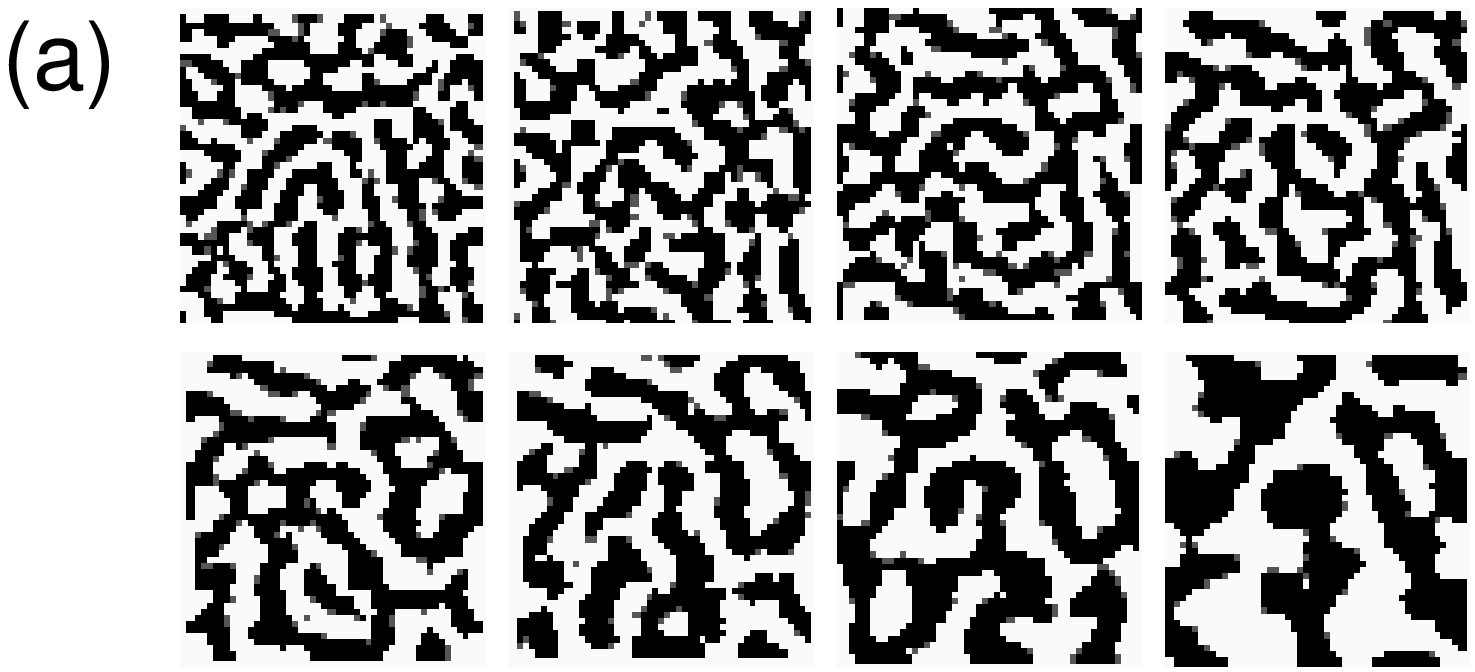}
\vspace*{0.5in}
\epsfxsize= 3in \epsfbox{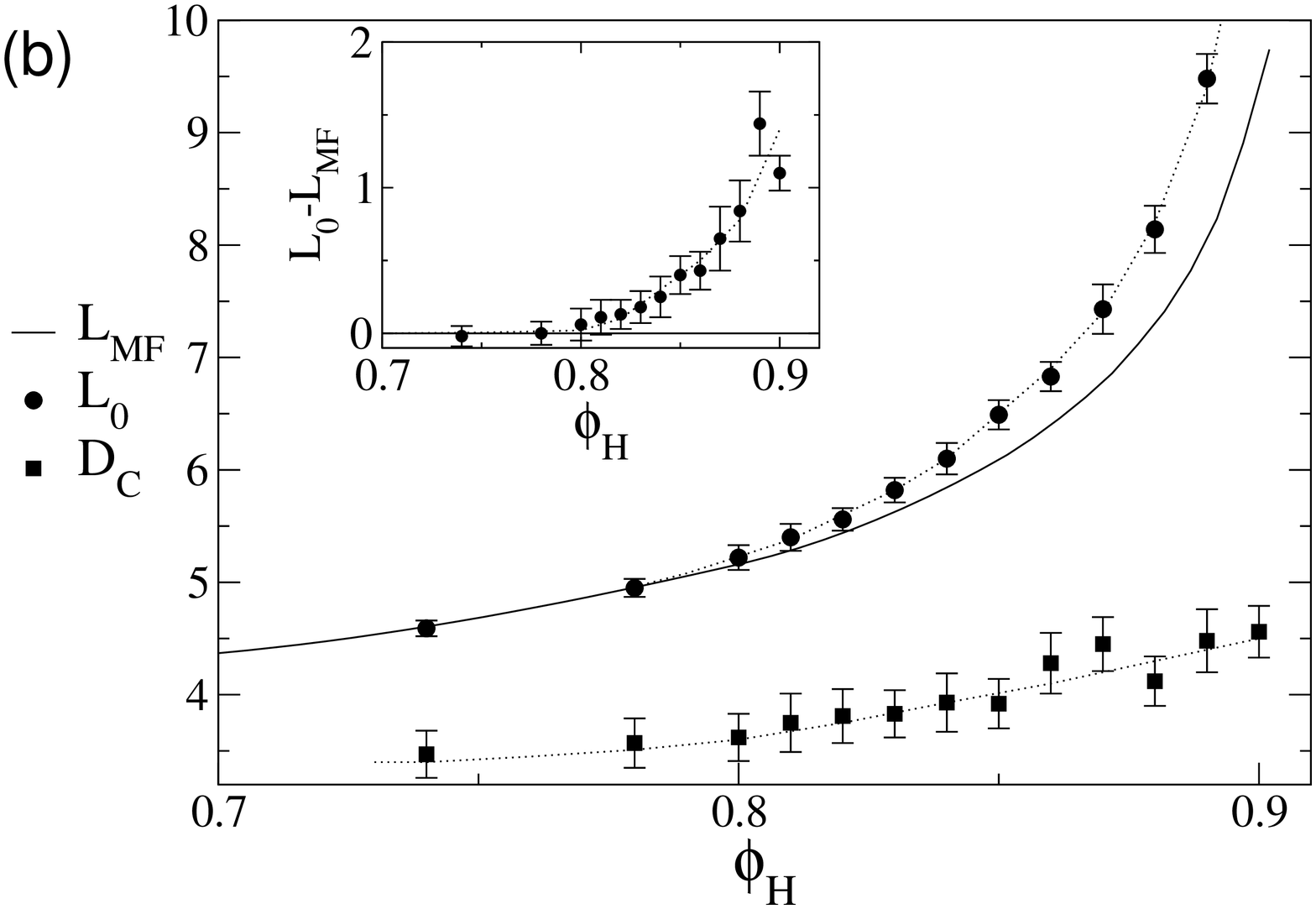}
\end{minipage}
\renewcommand
\baselinestretch{1.}
\caption{
Same as \fig~\ref{lengths125} for $\chi N = 12.$
As in \fig~\ref{lengths125}, the snapshots in (a) correspond to 
$\phi_H$ = 0.74, 0.78, 0.8, and 0.82 (first row from left), 
and $\phi_H$ = 0.84, 0.86, 0.88, and 0.9 (second row from left).
}
\renewcommand
\baselinestretch{1.5}
\label{lengths12}
\end{figure}

\begin{figure}[h]
\center
\begin{minipage}{2.5in}
\epsfxsize= 2.5in \epsfbox{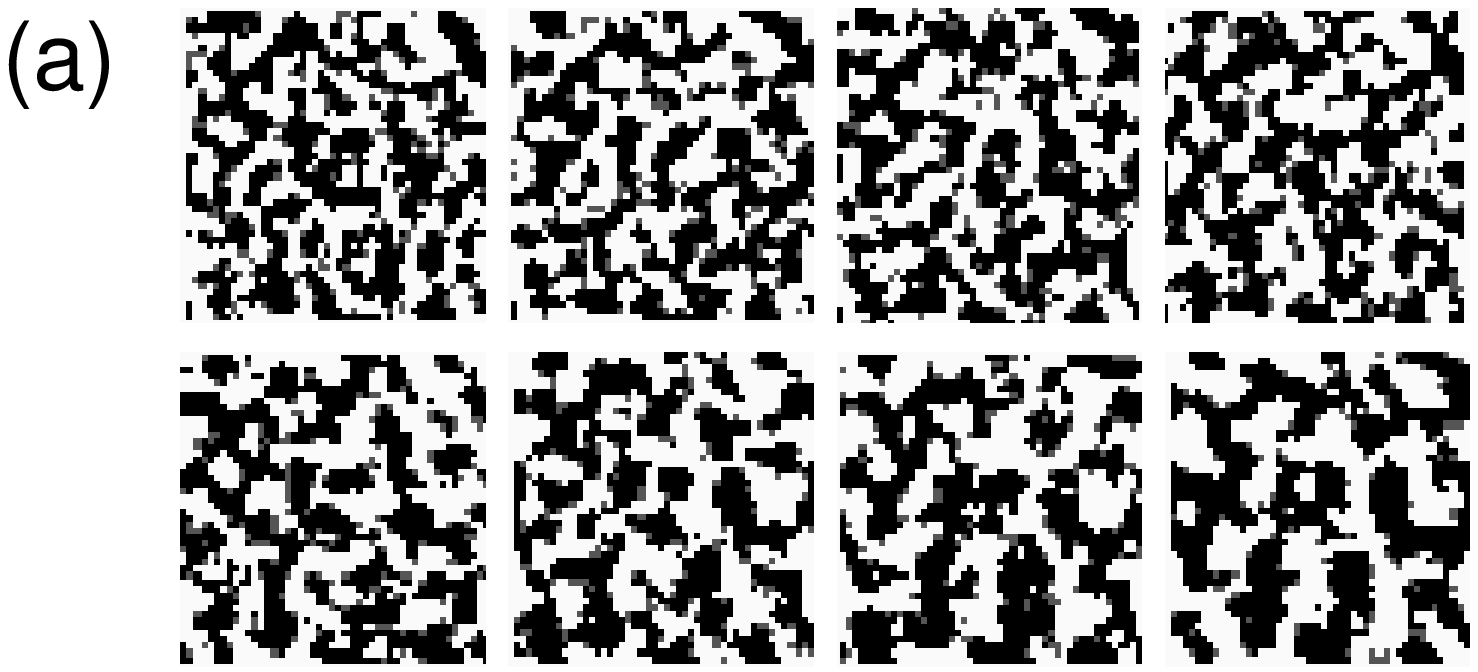}
\vspace*{0.5in}
\epsfxsize= 3in \epsfbox{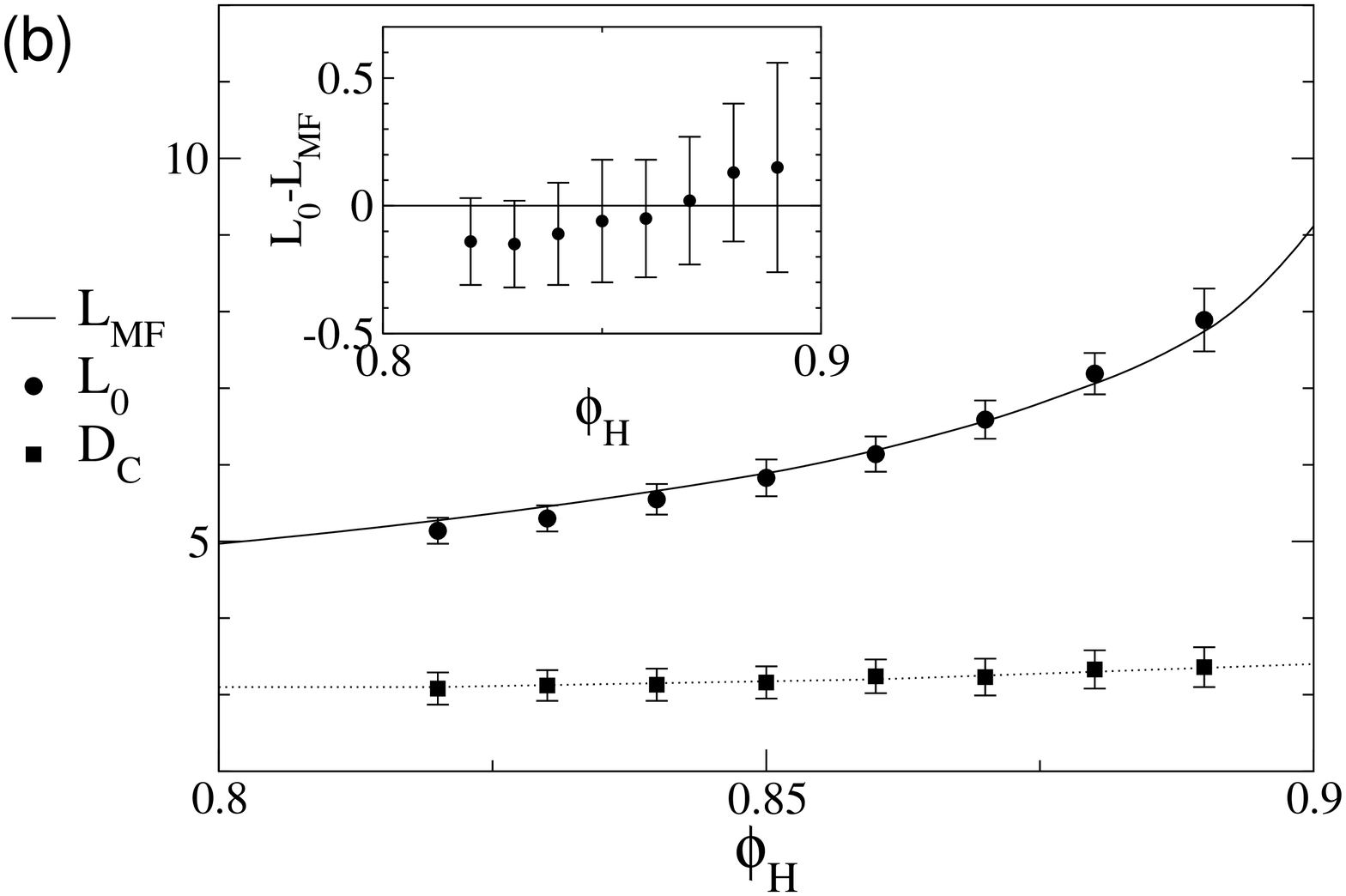}
\end{minipage}
\renewcommand
\baselinestretch{1.}
\caption{
Same as \fig~\ref{lengths125} for $\chi N = 11.5$
The snapshots in (a) correspond to 
$\phi_H$ = 0.82, 0.83, 0.84, and 0.85 (first row from left), 
and $\phi_H$ = 0.86, 0.87, 0.88, and 0.89 (second row from left).
}
\renewcommand
\baselinestretch{1.5}
\label{lengths115}
\end{figure}

In \figs~\ref{lengths125}, \ref{lengths12}, \ref{lengths115}, 
configurational snapshots\cite{footnote} 
and plots of equilibrated averages of $L_0$ and $D_C$ are displayed for various 
homopolymer volume fractions, $\phi_H$, at $\chi N$ = 12.5, 12, and 11.5. 
The deviation of $L_0$ from the corresponding mean-field values, $L_{MF}$, 
is shown in the inset. The Monte Carlo simulations were run up to approximately 
1.5 million Monte Carlo steps, including equilibration times of up to a 
few 100,000 steps. A Monte Carlo step includes one attempted random
local increment of $W_-$ per pixel, within ranges that were chosen such
that the Metropolis acceptance rate was 35 \%. As a general rule, the more 
disordered the configurations, the shorter was the equilibration phase. 
 
In all three cases, i.e., $\chi N$ = 12.5, 12, and 11.5, $D_C$ is in good
approximation constant over an extended $\phi_H$ range
(\figs~\ref{lengths125}, \ref{lengths12}, and \ref{lengths115}). This is plausible
because the curvature is induced by the copolymers, which are located
predominantly at the interfaces between microdomains. It should thus be
independent of the amount of homopolymers. Nevertheless, $D_C$ gets somewhat
smaller as $\chi N$ is decreased; the higher temperatures that correspond to
lower $\chi N$ facilitate the bending of the microdomain interfaces.

At $\chi N = 12.5$, coming from the lamellar phase at low $\phi_H$, the curvature
radius $D_C$ becomes comparable in size to the preferential length scale, $L_0$,
as well as the mean-field length scale, $L_{MF}$, around $\phi_H \approx 0.75$, 
which is when the lamellae begin to break up. This is in good
quantitative agreement with the location of the fluctuation-corrected
order-disordered transition of \fig~\ref{C50c}. At $\chi N$ = 12 and 11.5,
$D_C$ is well below both the mean-field periodicities, $L_{MF}$, and the
preferential length scales, $L_0$, in the $\phi_H$ range examined: these
configurations are in the disordered phase. We have thus identified the
mechanism by which fluctuations generate the microemulsion.

The preferential length scale $L_0$ and the mean-field length scale $L_{MF}$ 
are identical within error bars at small homopolymer concentrations $\Phi_H$.
Upon increasing $\Phi_H$ at $\chi N$ = 12.5 and 12, one notes that $L_0$ begins 
to deviate from $L_{MF}$ at a certain $\phi_H$. This signals the departure 
from defect-driven behavior and the onset of a more genuine morphology 
within the microemulsion phase. The effect is stronger for $\chi N = 12.5$ than 
for $\chi N = 12$. At $\chi N = 11.5$, it is barely noticeable anymore, 
and $L_0$ coincides within error bars with $L_{MF}$. However, the higher $\chi N$, 
the more smeared out $L_0$ becomes.

As pointed out already, a microemulsion is characterized by the existence of 
a preferential length scale within the disordered phase\cite{teubner,gompper}. 
In this sense, all configurations displayed in \figs~\ref{lengths125}, 
\ref{lengths12}, and \ref{lengths115} are microemulsions. 
The previous result, however, suggests a more diversified
classification into (a) a defect-driven region and (b) a ``genuine''
microemulsion. We shall denote them $D\mu E$ and $G\mu E$, respectively. 
In the first region, the microemulsion retains the characteristic 
length scale of a fluctuation-free lamellar phase, and the main effect of 
the fluctuations is to introduce lamellar defects, which destroy the lamellar 
order. Typical configurations are the first snapshots (top row)
in \figs~\ref{lengths125} and \ref{lengths12} a). In the second region,
the interfaces themselves fluctuate, as can be observed in the images
in the bottom row of \figs~\ref{lengths125} and \ref{lengths12} a).
This causes an effective increase of the characteristic length scale 
(or decrease in the characteristic wavevector). Deeper in the 
disordered phase, at low $\chi N$, the distinction between lamellar
defects and interfacial fluctuations is less obvious, one rather observes 
general structured, but unspecific disorder. In view of the fact 
that the characteristic wavevector coincides with the mean-field 
periodicity or, in the disordered phase, $q^*_{MF}$ (\fig~\ref{logSq}),
we classify it as $D\mu E$.

\section{SUMMARY AND CONCLUSION}

\begin{figure}[h]
\center
\begin{minipage}{3in}
\epsfxsize= 3in \epsfbox{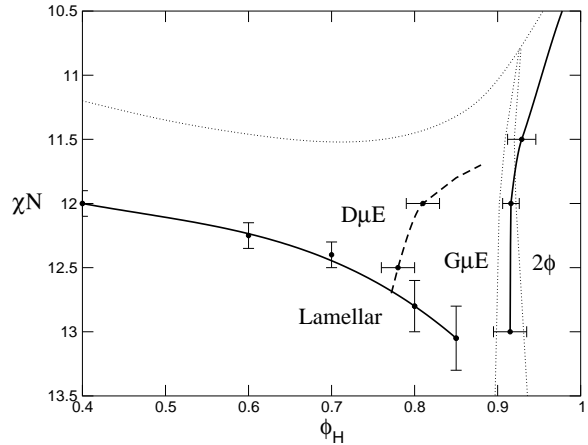}
\end{minipage}
\renewcommand
\baselinestretch{1.}
\caption{Substructure of the microemulsion. $D\mu E$: defect-driven, $G\mu E$:
genuine microemulsion morphology, $2\phi$: phase-separated region.
The solid lines show the fluctuation-corrected phase diagram of \fig~\ref{C50c},
the dotted lines the mean-field phase diagram, and the dashed line is just 
a guide for the eye.}
\renewcommand
\baselinestretch{1.5}
\label{C50d}
\end{figure}

In this paper, we have investigated the structure of the disordered
microemulsion phase in two-dimensional symmetrical ternary AB+A+B 
polymeric blends in different regions of the phase diagram. 
Not surprisingly, the structure of the microemulsion is strongest
at parameters $\chi N$ and $\phi_H$ where the mean-field approximation 
would predict an ordered phase, \ie where the disorder is brought about 
solely by fluctuations. We have corroborated our earlier result\cite{duechs}
regarding the mechanism that underlies the formation of the microemulsion phase 
in ternary AB+A+B polymeric blends: the lamellae break up when the curvature 
diameter of the microdomain boundaries becomes comparable to the periodicity 
of the lamellar phase\cite{morse}. We have further shown that the preferential 
length scale in the system deviates from its mean-field equivalent in a parameter 
subspace of the microemulsion only, demonstrating that the microemulsion 
region is divided into a defect-driven and a ``genuine'' part. 
In the resulting redrawing (\fig~\ref{C50d}) of the phase diagram 
\fig~\ref{C50c}, we have indicated the approximate location of the 
conjectured transition between the two, as marked by the occurrence 
of nonzero values for $L_0-L_{MF}$ in \fig~\ref{lengths125} 
and \ref{lengths12}. Comparing \fig~\ref{C50d} with the various
mean-field Lifshitz lines indicated in \fig~\ref{C50c}, we find that 
both the total monomer Lifshitz line and the homopolymer-homopolymer
Lifshitz line are far from the region of interest here. The numerical results
presented in \fig~\ref{logSq} confirm that this is still true for the
true (fluctuation-corrected) total monomer Lifshitz line.
We have not investigated the homopolymer-homopolymer structure factor. 
Since the short homopolymers in our system ($\alpha = 0.2$) swell the copolymers, 
the homopolymer-homopolymer Lifshitz line is not very far from the total monomer 
Lifshitz line.

It is possible that the $G\mu E$ phase coincides with the ``microemulsion'' 
($\mu E$) phase indicated in the experimental phase diagram of Schwahn
\etal \cite{schwahn}. Strictly speaking, however, the two systems cannot be 
compared directly to each other, because our simulations were carried out in two
dimensions only. We hope that three-dimensional calculations
will become feasible in the future.

We thank M. Matsen, V. Ganesan, and G. Fredrickson for fruitful discussions. 
This work was supported by the Deutsche Forschungsgemeinschaft (Germany).
The simulations were carried out on the CRAY T3E of the NIC institute 
in J\"ulich.

\appendix
\section{Graphical analysis}
In order to analyze configuration snapshots, it is often advantageous to
take the $\bar\phi_A$ distribution and convert its continuous values (between 0
and 1) to black-and-white bitmaps with white pixels for
$0\le\bar\phi_A(x,y)<0.5$, and black pixels for $0.5\le\bar\phi_A(x,y)\le 1$.
From these images, we can then extract the following parameters.

\subsection{Algorithm to calculate the circumference, $L_c$}
\begin{figure}[h]
\center
\begin{minipage}{2.5in}
\epsfxsize= 2.5in \epsfbox{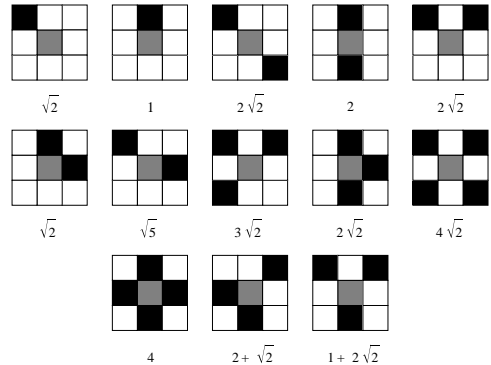}
\end{minipage}
\renewcommand
\baselinestretch{1.}
\caption{Local contributions to the circumference. The current (black) pixel
from step (3) here is painted grey. All diagrams are modulo $\pi/2$ rotations.}
\renewcommand
\baselinestretch{1.5}
\label{Ualgo}
\end{figure}

\begin{itemize}
\item[(1)] Besides the original bitmap, $B_1$, define another bitmap, $B_2$,
and paint it white, i.e., set all pixels to 0.
\item[(2)] Copy all border pixels from $B_1$ to $B_2$. A border pixel is
defined as one which is black (i.e., has a value of 1) and whose four nearest
neighbors have at least one white pixel among them.
\item[(3)] Scan through $B_2$ from the upper left to the lower right corner and
for each pixel check whether it is black. If so, proceed as follows:
\begin{itemize}
\item[(3a)] If any pair of adjacent nearest-neighbor (NN) and
next-nearest-neighbor (NNN) pixels of the current pixel are both black, paint the NN pixel
white. This is done to prevent ambiguities in the subsequent steps. In
practice, however, this case is quite rare and barely changes the end
result.
\item[(3b)] Check the distribution of black pixels among the nearest and
next-nearest neighbors according to \fig~\ref{Ualgo} and add the specified
numbers to the circumference.
\end{itemize}
\item[(4)] Once the entire lattice has been scanned, divide the result by 2 to
account for double counting in step (3b).
\end{itemize}

\subsection{Algorithm to calculate the curvature diameter, $D_C$}
The curvature diameter of the (combined) circumference, $L_c$, of
all black areas is defined as
\begin{equation}
D_C := 2\left[\frac{1}{L_c}\int\mbox{d}s~ \left|\frac{\mbox{d}{\bf t}}{\mbox{d}s}\right|^2\right]^{-\frac{1}{2}},
\end{equation}
with ${\bf t}$ the local (normalized) tangent vector. To calculate the quantity
\begin{equation}
\frac{1}{L_c}\int\mbox{d}s~ \left|\frac{\mbox{d}{\bf t}}{\mbox{d}s}\right|^2, 
\end{equation}
we pursue the following strategy:  
\begin{itemize}
\item[(I.1)] Make a border pixel bitmap as above.
\item[(I.2)] Scan the bitmap and stop when a black pixel is found.
\item[(I.3)] Start a data array, $(x^i_j,y^i_j)$, for the coordinates (numbered
$j\ge 0$) of a new line, $i\ge 0$. 
\item[(I.4)] Look for a (nearest or next-nearest) neighbor of the current
pixel. In most cases, there will be only one possibility to continue the
line. Otherwise, an arbitrary choice is made.
\item[(I.5)] Remove the current pixel from the bitmap. Make the neighboring
pixel from (4) the current pixel.
\item[(I.6)] Repeat (I.4-5) until no neighbor is found for the current
pixel. Then terminate that line.
\item[(I.7)] Repeat (I.2-6) until the bitmap has been cleared.\\
\item[(II.1)] For each line $i$, $(x^i_j,y^i_j)\equiv(x_j,y_j)$, as determined
in (I.1-7) that is longer than 3 points, set $j=1$.
\item[(II.2)] Calculate $|\mbox{d}{\bf t}/\mbox{d}s|^2$ according to
\begin{equation}
\left|\frac{\mbox{d}{\bf t}}{\mbox{d}s}\right|^2 = \frac{\mbox{d}{\bf t}_x^2 + \mbox{d}{\bf t}_y^2}{\mbox{d}s^2},
\end{equation}
\begin{eqnarray}
\mbox{d}{\bf t}_x &=& \frac{x_{j+2}-x_j}{\sqrt{(x_{j+2}-x_j)^2+(y_{j+2}-y_j)^2}} 
\\ &&
- \frac{x_{j+1}-x_{j-1}}{\sqrt{(x_{j+1}-x_{j-1})^2+(y_{j+1}-y_{j-1})^2}},
\nonumber \\
\mbox{d}{\bf t}_y &=& \frac{y_{j+2}-y_j}{\sqrt{(x_{j+2}-x_j)^2+(y_{j+2}-y_j)^2}} 
\\ &&
- \frac{y_{j+1}-y_{j-1}}{\sqrt{(x_{j+1}-x_{j-1})^2+(y_{j+1}-y_{j-1})^2}},
\nonumber
\end{eqnarray}
\begin{equation}
\mbox{d}x = \frac{x_{j+2}-x_{j+1}+x_j-x_{j-1}}{2},
\end{equation}
\begin{equation}
\mbox{d}y = \frac{y_{j+2}-y_{j+1}+y_j-y_{j-1}}{2},
\end{equation}
\begin{equation}
\mbox{d}s = \sqrt{\mbox{d}x^2 + \mbox{d}y^2}.
\end{equation}
\item[(II.3)] Increase $j$. Repeat (II.2-3) until the end of the line is
reached.
\item[(II.4)] Repeat (II.1-3) for all lines.
\item[(II.5)] Apply
\begin{equation}
L_c = \int\mbox{d}s.
\end{equation}
\end{itemize}
Note that this algorithm produces a slightly different (smaller) value for
$L_c$ than the one presented in the last section. This because the length of
each line is effectively truncated by three points. $L_c$ here acts only as a
normalizing factor and for consistency should be calculated as indicated. If
interested in $L_c$ itself, one should use the algorithm of the previous
section.


\begin{thebibliography}{99}
\bibitem{footnote} The attentive reader will notice a certain similarity of 
some configurations, especially those at higher $\chi N$. The corresponding 
simulations were indeed started independently of each other, but with the 
same seed number for the random number generator (which one is allowed to do).
\end{thebibliography}
\end{document}